\documentclass[12pt]{article}
\usepackage{amsfonts}
\usepackage{amssymb}
\usepackage{graphics,amsmath}
\usepackage{color}

\def\hybrid{\topmargin -20pt    \oddsidemargin 0pt
        \headheight 0pt \headsep 0pt
        \textwidth 6.35in       % BS paper
        \textheight 9.25in       % BS paper
        \marginparwidth .875in
        \parskip 5pt plus 1pt   \jot = 1.5ex}

\hybrid

\def\baselinestretch{1.2}

\catcode`\@=11

\def\marginnote#1{}
\newcount\hour
\newcount\minute
\newtoks\amorpm
\hour=\time\divide\hour by60
\minute=\time{\multiply\hour by60 \global\advance\minute by-\hour}
\edef\standardtime{{\ifnum\hour<12 \global\amorpm={am}%
        \else\global\amorpm={pm}\advance\hour by-12 \fi
        \ifnum\hour=0 \hour=12 \fi
        \number\hour:\ifnum\minute<10 0\fi\number\minute\the\amorpm}}
\edef\militarytime{\number\hour:\ifnum\minute<10 0\fi\number\minute}
\def\draftlabel#1{{\@bsphack\if@filesw {\let\thepage\relax
   \xdef\@gtempa{\write\@auxout{\string
      \newlabel{#1}{{\@currentlabel}{\thepage}}}}}\@gtempa
   \if@nobreak \ifvmode\nobreak\fi\fi\fi\@esphack}
        \gdef\@eqnlabel{#1}}
\def\@eqnlabel{}
\def\@vacuum{}
\def\draftmarginnote#1{\marginpar{\raggedright\scriptsize\tt#1}}

\def\draft{\oddsidemargin -.5truein
        \def\@oddfoot{\sl preliminary draft \hfil
        \rm\thepage\hfil\sl\today\quad\militarytime}
        \let\@evenfoot\@oddfoot \overfullrule 3pt
        \let\label=\draftlabel
        \let\marginnote=\draftmarginnote
   \def\@eqnnum{(\theequation)\rlap{\kern\marginparsep\tt\@eqnlabel}%
\global\let\@eqnlabel\@vacuum}  }

\def\preprint{\twocolumn\sloppy\flushbottom\parindent 2em
        \leftmargini 2em\leftmarginv .5em\leftmarginvi .5em
        \oddsidemargin -.5in    \evensidemargin -.5in
        \columnsep .4in \footheight 0pt
        \textwidth 10.in        \topmargin  -.4in
        \headheight 12pt \topskip .4in
        \textheight 6.9in \footskip 0pt
        \def\@oddhead{\thepage\hfil\addtocounter{page}{1}\thepage}
        \let\@evenhead\@oddhead \def\@oddfoot{} \def\@evenfoot{} }

\def\numberbysection{\@addtoreset{equation}{section}
        \def\theequation{\thesection.\arabic{equation}}}

\def\underline#1{\relax\ifmmode\@@underline#1\else
        $\@@underline{\hbox{#1}}$\relax\fi}

\def\titlepage{\@restonecolfalse\if@twocolumn\@restonecoltrue\onecolumn
     \else \newpage \fi \thispagestyle{empty}\c@page\z@
        \def\thefootnote{\fnsymbol{footnote}} }

\def\endtitlepage{\if@restonecol\twocolumn \else \newpage \fi
        \def\thefootnote{\arabic{footnote}}
        \setcounter{footnote}{0}}  %\c@footnote\z@ }

\catcode`@=12
\relax

\def\figcap{\section*{Figure Captions\markboth
        {FIGURECAPTIONS}{FIGURECAPTIONS}}\list
        {Figure \arabic{enumi}:\hfill}{\settowidth\labelwidth{Figure
999:}
        \leftmargin\labelwidth
        \advance\leftmargin\labelsep\usecounter{enumi}}}
 \relax
\def\tablecap{\section*{Table Captions\markboth
        {TABLECAPTIONS}{TABLECAPTIONS}}\list
        {Table \arabic{enumi}:\hfill}{\settowidth\labelwidth{Table
999:}
        \leftmargin\labelwidth
        \advance\leftmargin\labelsep\usecounter{enumi}}}
 \relax
\def\reflist{\section*{References\markboth
        {REFLIST}{REFLIST}}\list
        {[\arabic{enumi}]\hfill}{\settowidth\labelwidth{[999]}
        \leftmargin\labelwidth
        \advance\leftmargin\labelsep\usecounter{enumi}}}
 \relax
\makeatletter
\newcounter{pubctr}
\def\publist{\@ifnextchar[{\@publist}{\@@publist}}
\def\@publist[#1]{\list
        {[\arabic{pubctr}]\hfill}{\settowidth\labelwidth{[999]}
        \leftmargin\labelwidth
        \advance\leftmargin\labelsep
        \@nmbrlisttrue\def\@listctr{pubctr}
        \setcounter{pubctr}{#1}\addtocounter{pubctr}{-1}}}
\def\@@publist{\list
        {[\arabic{pubctr}]\hfill}{\settowidth\labelwidth{[999]}
        \leftmargin\labelwidth
        \advance\leftmargin\labelsep
        \@nmbrlisttrue\def\@listctr{pubctr}}}
 \relax
\makeatother
\newskip\humongous \humongous=0pt plus 1000pt minus 1000pt

\newif\ifdtup

\relax

\def\be{\begin{equation}}
\def\ee{\end{equation}}
\def\ba{\begin{eqnarray}}
\def\ea{\end{eqnarray}}

\def\a{\alpha}

\def\no{\noindent}

\def\IR{\relax{\rm I\kern-.18em R}}
\def\II{\relax{\rm 1\kern-.35em1}}

\renewcommand{\theequation}{\thesection.\arabic{equation}}
\csname @addtoreset\endcsname{equation}{section}

\def\IR{\relax{\rm I\kern-.18em R}}
\def\inv{^{\raise.15ex\hbox{${\scriptscriptstyle -}$}\kern-.05em 1}}

\begin{document}

\begin{titlepage}
\begin{center}

\vskip .5in

{\LARGE Quantization of Galilean Electrodynamics:\\a non-trivially trivial theory}
\vskip 0.4in

{\bf Rafael Hern\'andez}$^{\dag}$,  \phantom{x} {\bf Juan Miguel Nieto Garc\'ia}$^{\ddag}$\phantom{x} and \phantom{x} {\bf Ander Urtiaga}$^{\perp}$
\vskip 0.1in

$^{\dag}$ Departamento de F\'{\i}sica Te\'orica \\
and \\
Instituto de F\'isica de Part\'iculas y del Cosmos (IPARCOS),\\
Universidad Complutense de Madrid \\
$28040$ Madrid, Spain \\
{\footnotesize{\tt rafael.hernandez@fis.ucm.es}}

\vskip 0.2in

$\ddag$ Departamento de Matem\'atica Aplicada a las Tecnolog\'{\i}as de la Informaci\'on \\ y las Comunicaciones, ETSIS de Telecomunicaci\'on, \\ 
Universidad Polit\'ecnica de Madrid \\
C. Nikola Tesla s/n \\
$28031$ Madrid, Spain \\
{\footnotesize{\tt juanmiguel.nietogarcia@upm.es}}

\vskip 0.2in

$\perp$ Departamento de F\'{\i}sica \\\ 
Universidad del Pa\'{\i}s Vasco, UPV/EHU \\
$48080$ Bilbao, Spain \\
{\footnotesize{\tt ander.urtiaga@ehu.eus}}

\end{center}

\vskip .4in

\centerline{\bf Abstract}
\vskip .1in
\no
\noindent
We consider the quantization of the non-relativistic limit of electrodynamics called Galilean electrodynamics. 
To that end, we apply the Dirac bracket formalism to understand the constraints and dynamics of the theory 
and analyze its gauge structure. We show that the theory is fully constrained, 
which eliminates all the degrees of freedom from the path integral. 

\vskip .4in
\noindent

\end{titlepage}

\vfill
\eject

\def\baselinestretch{1.2}

%%%%%%%%%%%%%%%%%%%%%%%%%%%%%%%%%%%%%%%%%%%%%%%%%%%

\baselineskip 20pt

%%%%%%%%%%%%%%%%%%%%%%%%%%%%%%%%%%%%%%%%%%%%%%%%%%%%%%%%%%%%%%%%%%%%%%%%
%%%%%%%%%%%%%%%%%%%%%%%%%%%%%%%%%%%%%%%%%%%%%%%%%%%%%%%%%%%%%%%%%%%%%%%%

\section{Introduction}

Interest in Galilean and Carrollian theories has been growing in the last few years. Although it is well-known that theories of nature have to be Lorentz-invariant, 
models with non-Lorentzian symmetry could be useful to describe certain regimes, like many-body problems in condensed matter, e.g. to understand the Quantum Hall Effect \cite{Son}, 
or the near-horizon limit of a non-extremal Killing horizon \cite{Carroll,Fontanella2022}. 
In the context of string theory, there has also been an increasing interest in studying the propagation of strings in Newton-Cartan backgrounds, 
which can be used to describe both non-relativistic and Carrollian theories. 
This has also led to the question whether the AdS/CFT correspondence survives after applying either the non-relativistic or the Carrollian limit. 
Recent works suggest that the answer to this question is affirmative in both cases \cite{Lambert:2024uue}-\cite{
%Fontanella:2024rvn,Lambert:2024yjk,Fontanella:2024kyl,Fontanella:2025tbs,
Fontanella:2026gaq}. 
As the proposed dual to non-relativistic strings propagating in AdS$_5\times $S$^5$ 
is the non-relativistic limit of four-dimensional Yang-Mills with $\mathcal{N}=4$ supersymmetry, usually denoted as Galilean Yang-Mills, it is worth revisiting some known results 
and performing new computations in this theory.

In this article, we will study the quantization of the abelian version of Galilean Yang-Mills theory, also known as Galilean electrodynamics (and usually shortened to GED). The action for Galilean electrodynamics was constructed in~\cite{Santos} as the action that provides the non-relativistic limit of Maxwell's equations, first derived in~\cite{ged}. The reason why the action was not constructed soon after the equations 
of motion were proposed is because the limit is subtle, see e.g. the discussion regarding the Helmholtz conditions in \cite{Banerjee:2019axy}. Taking the non-relativistic limit at the level of Maxwell's action gives us a different result than taking the limit 
at the level of the equations of motion, and auxiliary fields are required for these two computations to match. Galilean electrodynamics has been more extensively analyzed in recent literature. 
Symmetries were calculated in~\cite{Festuccia:2016caf,Bagchi:2022twx}, the analysis of the canonical structure was performed in \cite{Banerjee:2019axy}, correlation functions were studied in~\cite{Bagchi:2014ysa}, 
renormalization was studied in~\cite{Chapman:2020vtn} for 2+1 dimensions and a path integral formulation of the theory was performed in~\cite{Banerjee:2022uqj} for 3+1 dimensions. 
Nevertheless, recently it has been shown that the classification of symmetries from~\cite{Festuccia:2016caf,Bagchi:2022twx} were incomplete, and a more exhaustive result 
was given in~\cite{Fontanella:2024hgv}. These results also open the door to an analysis of the theory from a perspective more akin to Conformal Filed Theories, see \cite{Fontanella:Future}. 
Motivated by these developments, and by the fact that the quantization of this theory has not yet been discussed in sufficient detail, 
in this article we perform an analysis of the gauge structure and constraints of this theory with the purpose of computing its path integral.

The remaining part of this article is organized as follows. In section 2 we will present a brief summary of the null reduction method and the action of Galilean electrodynamics. 
In section 3 we will apply the Dirac bracket formalism to study and classify the dynamical structure of Galilean electrodynamics. We will also discuss the gauge fixing procedure 
and analyse the classical equations of motion. In section 4 we will compute the path integral of Galilean electrodynamics and show that it only contains zero modes. 
In section 5 we will conclude with a brief summary of our results, together with some additional remarks and a discussion on future directions of research.

%%%%%%%%%%%%%%%%%%%%%%%%%%%%%%%%%%%%%%%%%%%%%%%%%%%%%%%%%%%%%%%%%%%%%%%%%%%%%%%%%%%%%%%%%%%%%%%%%%%%
%%%%%%%%%%%%%%%%%%%%%%%%%%%%%%%%%%%%%%%%%%%%%%%%%%%%%%%%%%%%%%%%%%%%%%%%%%%%%%%%%%%%%%%%%%%%%%%%%%%%

\section{Galilean electrodynamics}

In this section, we will present the action of Galilean electrodynamics. Naively taking the Galilean limit of Maxwell's action and of Maxwell's equations leads to inconsistent results, 
because the equations that are obtained are not the Euler-Lagrange equations of the non-relativistic action. To find the action that provides the correct Euler-Lagrange equations 
we can either take the non-relativistic limit of Maxwell's action together with a free scalar field~\cite{Santos,Festuccia:2016caf,Berg}, 
or we can consider instead Maxwell's action in one additional dimension and perform dimensional reduction on a null direction~\cite{5dimensionalmetric}. In this section we will present 
the null reduction method. In order to proceed, we will note first that both four-dimensional Lorentzian and Galilean spacetimes can be described in terms of a five-dimensional Lorentz metric, 
with a covariant constant scalar field that vanishes when the Galilean limit is performed~\cite{5dimensionalmetric}. We will choose light-cone coordinates because 
these are the ones that reduce to Galilean invariance~\cite{Galileanembebimiento}-\cite{OKTO}. In particular, we consider the following metric
\be
    g_{\mu\nu}=\begin{pmatrix}
        \mathbb{I}_{3\times 3} & 0 & 0\\
0 & 0 & -1\\
0 & -1 & 0
    \end{pmatrix} \ ,
    \label{galileanmetric}
\ee
which defines the following scalar product between two five-vectors, 
\be
\label{scalarproductgalileanmanifold} x^\mu y_\mu=x^{a} y_{a} -x^4 y^5-x^5 y^4 \ ,
\ee
where $\mu$ runs from $1$ to $5$, while latin indices run over spatial directions, $a=1,2,3$. The scalar product is invariant under the transformations
\be
    \label{Galboost} {x}'^{a} = {x}^{a} - v^{a}x^4 \ ,\quad x'^4=x^4 \ ,\quad x'^5=x^5 - {v}^{a} {x}_{a} + \frac{1}{2} v^{a} v_{a} x^4 \ ,
\ee
which are a generalization of the Galilean transformation rules, where ${v}^{a}$ are the components of the relative velocity. In this way, four-dimensional Galilean spacetime is embedded 
in a five-dimensional spacetime and the position five-vector is given by 
\be
    x^\mu=\left(x^a,t,s\right),
\ee
that allows to identify the five-momentum as
\be
p_\mu = - i\partial_\mu=(-i \partial_a,-i \partial_t, - i \partial_s)=(p_a,-E,-m) \ ,
\ee
where $p^4=-p_5=m$ is the mass, and $p^5=-p_4=E$ is the energy. 
We will now follow \cite{Santos} and write the general action functional for $N$ different fields on five-dimensional Galilean spacetime as 
\be
S\left[\Phi\right]=\int d^5x \, \mathcal{L}\left[\Phi_\alpha(x^\mu),\partial_\mu\Phi_\alpha(x^\mu)\right],
\ee
where the extra coordinate $x^{5}$ is defined over the real numbers. In this way, integrals along the $x^{5}$ direction can be interpreted as
\be
\int dx^5\longrightarrow \lim\limits_{l\rightarrow\infty}\frac{1}{2l}\int_{-l}^l ds \ .
\ee
Therefore, if the integral over the manifold is independent of $s$, the integration reduces to the usual one in four-dimensional spacetime.

Let us now apply this construction to the Lagrangian for five-dimensional electrodynamics, 
\be
\mathcal{L}_{5d}=- \frac{1}{4}F_{\mu\nu}F^{\mu\nu}\ ,\label{LagrangianoGED}
\ee
with the gauge field split as 
\be
A^\mu(\vec{x},t)= (A^a,\phi_e,\phi_m)\ ,\hspace{5mm}A_\mu(\vec{x},t)=(A_a,-\phi_m,-\phi_e) \ . \label{split}
\ee
In this convention, the constraints $A^5=0$ and $A^4=0$ define, respectively, the electric and magnetic regimes of the theory. As we want our photons to be massless, we need $p_s=-m=0$. 
This imposes $A_\mu$ to be independent of $s$, i.e. that $\partial_s A_\mu=\partial^t A_\mu=0$. We should also note that 
the invariance of the Faraday tensor under gauge transformations,  
\be
A_\mu\hspace{2mm}\longrightarrow\hspace{2mm}A_\mu+\partial_\mu\lambda(\vec{x},t) \ ,
\label{transformgauge}
\ee
implies that transformations of $A_5$ are limited to a change by a constant amount, 
because $A_\mu$ does not depend on $s$. 
Substituting the metric  (\ref{galileanmetric}) and with the notation in (\ref{split}), the Lagrangian \eqref{LagrangianoGED} takes the following form 
\be
\mathcal{L}_{GED}= -\frac{1}{4} F_{ab} F^{ab} + (\partial_a \phi_m + \partial_t A_a) \partial_a \phi_e +\frac{1}{2} (\partial_t \phi_e)^2 \ , \label{LGED}
\ee
where we have imposed that all the fields are independent of $x^5=s$. The canonical momenta associated to this Lagrangian are 
\be
\pi_\mu=\frac{\partial\mathcal{L}}{\partial(\partial_t A^\mu)}=-F^{4}\null_\mu=-(\partial^t A_\mu-\partial_\mu A^4)=\partial_\mu A^4 \ ,  
\ee
where we have used that $\partial^t=-\partial_s$. Recalling the notation in (\ref{split}), they can be written as    
\be
\pi_a
    =\partial_a \phi_e \ , \qquad \pi_4=\partial_t\phi_e \ ,\qquad \pi_5=0 \ ,
\label{pi5}
\ee
and thus the Hamiltonian is given by 
\be
H_{GED}=\int d^3x\left[ \frac{1}{2} \pi_4^2+ \frac{1}{4} (F_{ab})^2-\partial_a \phi_m \partial_a \phi_e \right] \ . \label{HGED}
\ee
Notice that we can use integration by parts to write $\phi_m$ as a Lagrange multiplier for the constraint $\partial_a \pi_a \equiv \vec{\nabla} \cdot \vec{E}=0$, 
as it happens in conventional electrodynamics. We should also mention that the Legendre transformation used to construct this Hamiltonian is singular 
because the matrix $T_{\mu \nu}=\frac{\delta \pi_\mu}{\delta A^\nu}$ is not invertible. Consequently, the theory is constrained.

%%%%%%%%%%%%%%%%%%%%%%%%%%%%%%%%%%%%%%%%%%%%%%%%%%%%%%%%%%%%%%%%%%%%%%%%%%%%%%%%%%%%%%%%%%%%%%%%%%%%
%%%%%%%%%%%%%%%%%%%%%%%%%%%%%%%%%%%%%%%%%%%%%%%%%%%%%%%%%%%%%%%%%%%%%%%%%%%%%%%%%%%%%%%%%%%%%%%%%%%%

\section{Dynamical structure of Galilean electrodynamics} 

Canonical quantization, which is performed by transforming the Poisson bracket into a commutator, cannot be directly applied to Galilean electrodynamics because it is a constrained system. 
This can be easily understood from 
the Euler-Lagrange equations of the five-dimensional theory before performing the null reduction, 
\be
\left[g_{\mu\nu}(\partial_\rho\partial^\rho)-\partial_\mu\partial_\nu\right] A^\nu=0 \ .
\ee
These equations are non-invertible, which implies an excess of degrees of freedom that we will have to handle with a gauge condition in order to quantize 
the theory. In fact, it is clear from the analysis in the previous section that the system has the following four primary constraints
\begin{equation}
    \chi_1 = \pi_5 =0 \ , \qquad \chi_{2,a} = \pi_a- \partial_a \phi_e \ . \label{primaryconst}
\end{equation}
In this section, we will employ the Dirac bracket formalism to the study of Galilean electrodynamics as a constrained system~\cite{dirac1964lectures}.
We will start with an abridged review of the method. 

\subsection{Dirac bracket formalism}

Naively quantising a constrained system by identifying the commutator with the Poisson bracket, defined as
\be
    \left\{F(\vec{x},t),G(\vec{y},t)\right\}_{PB}
    =\int d^3z \sum_{\mu=1}^5 \left(\frac{\delta F(\vec{x},t)}{\delta A_\mu(\vec{z},t)}\frac{\delta G(\vec{y},t)}{\delta \pi^\mu(\vec{z},t)}-\frac{\delta F(\vec{x},t)}{\delta \pi^\mu(\vec{z},t)}\frac{\delta G(\vec{y},t)}{\delta A_\mu(\vec{z},t)}\right) \ ,
\ee
where $F(\vec{x},t)$ and $G(\vec{y},t)$ and where the functional derivative is defined as
\be
    \frac{\delta A^\mu(\vec{x},t)}{\delta A^\nu(\vec{z},t)}=\frac{\delta \pi^\mu(\vec{x},t)}{\delta\pi^\nu(\vec{z},t)}
    =\delta^\mu_\nu \, \delta^{3}(\vec{x}-\vec{z})\ ,
\ee
leads to inconsistent results. This happens because the Poisson bracket does not automatically incorporate the constraints, so using it carelessly 
may lead to time evolution that takes the system out of the constraint surface. The aim of the Dirac bracket formalism is to modify the Poisson bracket 
in such a way that time evolution is consistent with the constraints. As a starting point, let us first recall that the Hamiltonian of a constrained system can be modified by adding terms proportional to the constraints,
\be
     H^\ast=H+\sum_s u_s \chi_s\approx H \ , \label{modifiedLagrangian}
\ee
where $\chi_s$ are the constraints, $u_s$ are some arbitrary functions, and where the symbol $\approx$ has been employed to indicate a weak equality, 
i.e. an equality that may not hold unless the constraints are imposed. Demanding that the constraints are fulfilled at any given time is equivalent to demanding 
that their time evolution with respect to $H^\ast$ weakly vanishes, that is
\be
    \dot{\chi}_m= \left\{\chi_m,H\right\}_{PB}+\sum_s u_s\left\{\chi_m,\chi_s\right\}_{PB} \approx 0 \ .\label{eqconsistencia}
\ee
These equations are called consistency equations, and they fall in one of the four following cases: inconsistent equalities, that imply that the chosen 
Lagrangian is not consistent; trivial identities, that provide no information; algebraic equations for the coefficients $u_s$, that can used to fix them; or a set of new constraints.
These additional constraints, obtained from consistency equations, are called secondary constraints, while the originals are named primary constraints. 
Once we have found all the secondary constraints, we add new terms proportional to these new constraints to the Hamiltonian and repeat the process once again. 
The procedure, usually called Dirac-Bergmann algorithm, is iterated until no more constraints are generated. Although this method produces a set of coefficients $u_s$ that solve 
the consistency equations (\ref{eqconsistencia}), the solutions may not be unique. This is the case, for example, in systems with gauge redundancy, 
like the one we are studying. Including gauge fixing conditions as further constraints into the above procedure may help to make the solution unique.

At this point, we should distinguish between first-class and second-class constraints. A constraint $\chi_r$ is said to be of first-class if it satisfies 
\be
      \left\{\chi_r,\chi_s \right\}_{PB}\approx 0 \qquad \forall s\ ,
\ee
that is, if it Poisson-commutes with any other constraint, either primary or secondary. Constraints that are not first-class are called second-class constraints. 
It is clear that second-class constraints are an obstruction to quantization. When quantizing the system, we need to set all the constraints to zero as operators. However, 
if two constraints do not Poisson-commute, their quantum counterparts would not commute and  we cannot simultaneously diagonalize them. 
Therefore, to consistently set these constraints to zero we need to perform canonical quantization on a generalization of the Poisson bracket where the second-class constraints commute. 
Such a generalization is the Dirac bracket, which is defined as 
\be
\left\{f,g\right\}_D=\left\{f,g\right\}_{PB}-\sum_{m,n} \left\{f,\tilde{\Phi}_m\right\}_{PB}M^{-1}_{mn}\left\{\tilde{\Phi}_n,g\right\}_{PB} \ ,\label{diracbracket}
\ee
where $\tilde{\Phi}_m$ are only second-class constraints and where the matrix $M_{mn}$, which is always invertible~\cite{Dirac_1950}, is defined as 
\be
    M_{mn}=\left\{ \tilde{\Phi}_m,\tilde{\Phi}_n\right\}_{PB} \ .
\ee
By construction, the Dirac bracket between two second-class constrains vanishes and it is thus the adequate structure to quantize the theory.

\subsection{Dirac bracket formalism applied to Galilean electrodynamics}
\label{constraints}

We will now apply the Dirac-Bergmann algorithm to the Lagrangian for Galilean electrodynamics, eq.~\eqref{LGED}.~\footnote{Although an analysis of the constraints of Galilean electrodynamics 
was presented in \cite{Banerjee:2019axy}, we believe that several points, such as the classification of the first-class and second-class constraints or the gauge fixing procedure, merit further analysis.} We will start by iteratively demanding that the constraints are conserved by time evolution. 
This will either give us additional constraints or fix the Lagrange multipliers $u_s$. Next, we will classify these constraints in order to find the number of degrees of freedom of our theory. We will start from the Hamiltonian
\begin{equation}
    H_1^\ast=\int d^3x\left[ \left(\frac{1}{2} \pi_4^2+ \frac{1}{4} (F_{ab})^2-(\partial_a \phi_m) (\partial_a \phi_e) \right) + u_1 \pi_5 + u_{2,a} (\pi_a - \partial_a \phi_e) \right] \ .
\end{equation}
The consistency conditions (\ref{eqconsistencia}), imposed for the primary constraints \eqref{primaryconst}, lead to
\begin{align}
    \left\{\chi_1,H_1^*\right\}_{PB}=&\left\{\chi_1,H_{GED}\right\}_{PB}+u_2\left\{\chi_1,\chi_2\right\}_{PB}=-\partial_a \partial_a \phi_e=-\chi_3 \ , \\
    \left\{\chi_{2,a},H_1^*\right\}_{PB}=&\left\{\chi_2,H_{GED}\right\}_{PB}+u_1\left\{\chi_2,\chi_1\right\}_{PB}=-(\partial_a \partial_b A_ b - \partial_b \partial_b A_a)-\partial_a \pi_4 \notag \\
    =& \:\: \partial_b F^{ba} - \partial_a \pi_4=-\chi_{4,a} \ .
\end{align}
The Hamiltonian $H_2^*$ is obtained by 
adding the secondary constraints $\chi_3$ and $\chi_{4,a}$ to the Hamiltonian $H_1^\ast$. Repeating the process we find
\begin{displaymath}
    H_2^\ast=\int d^3x\left[ \left(\frac{1}{2} \pi_4^2+ \frac{1}{4} (F_{ab})^2-(\partial_a \phi_m) (\partial_a \phi_e) \right) + u_1 \chi_1 + u_{2,a} \chi_{2,a}+ u_3 \chi_3+ u_{4,a} \chi_{4,a} \vphantom{\frac{1}{2} \pi_4^2}\right] \ ,
\end{displaymath}\vspace{-1cm}
\begin{align}
    \left\{\chi_1,H_2^*\right\}_{PB}&=\chi_3 \ , \label{DBP1}\\
    \left\{\chi_{2,a},H_2^*\right\}_{PB}&=\partial_b \partial_b u_{4,a}+\partial_b F^{ba} - \partial_a \pi_4=\partial_b \partial_b u_{4,a}-\chi_{4,a} \ , \\
    \left\{\chi_3,H_2^*\right\}_{PB}&= \partial_c \partial_c \pi_4 - \partial_b \partial_b \partial_c u_{4,c} \ , \\
    \left\{\chi_{4,a},H_2^*\right\}_{PB}&=-\partial_b \partial_b (\partial_a \phi_m + u_{2,a} + \partial_a u_3 ) \ . \label{DBP4}
\end{align}
From here we conclude that
\be
    u_{2,a}  =-(\partial_a \phi_m +\partial_a u_3 ) \ , \quad
    u_{4,a}  =-P^T_{ab} A^b + \hat{u}_a \ ,
\ee
where $\hat{u}_a$ fulfils $\partial_c \hat{u}_c=\pi_4$ and $P^T_{ab}$ is the projector on the transverse direction, formally given by
\begin{equation}
    P^T_{ab} = \delta_{ab}+\frac{\partial_a \partial_b}{\partial_c \partial_c} \ .
\end{equation}
Substituting these values and repeating the process again gives us
\begin{align}
    \{\chi_1, H_3^\ast\} &=\chi_3 \approx 0 \ , &  \{\chi_{2,a}, H_3^\ast\} &=P^T_{ab} \, \chi_{4,b}\approx 0 \ , &    \{\chi_{4,a}, H_3^\ast\} &= \{\chi_3, H_3^\ast\} =0 \ ,
\end{align}
where $H_3^\ast$ is $H_2^\ast$ after we have substituted the values for $u_2$ and $u_4$. As we can see, no further constraints appear and we have finished applying Dirac-Bergmann algorithm.

For future computations it is useful to further separate the constraint $\chi_{4,a}$ into its transverse and longitudinal part with the help of the projector $P^T_{ab}$. We will rewrite $\chi_{4,a}$ as
\begin{equation}
    \chi_{4,a}=\partial_a \pi_4 - \partial_b F^{ba}=\partial_a \pi_4 - \partial_b \partial_b P^T_{ac} A^c \ ,
\end{equation}
where the first term only contributes to the longitudinal part and the second term only contributes to the transverse part. Consequently, we can separate $\chi_{4,a}$ into
\begin{equation}
    \chi_{4}^L= \partial_a\partial_a \pi_4 \ , \qquad \text{ and } \qquad \chi_{4,a}^T=\partial_b \partial_b P^T_{ac} A^c \ . \label{splitconstraint}
\end{equation}

A naive counting of the constraints provides one first-class constraint, $\chi_1$, and seven second-class constraints, as can be infered from equations (\ref{DBP1})-(\ref{DBP4}). 
As first-class constraints fix two degrees of freedom while second-class constraints fix one degree of freedom~\cite{Thiemann}, 
and we started with a ten-dimensional phase space, the dimension of the constrained phase space is reduced to one, 
 $   10-(2\cdot 1)-(1\cdot 7)=1$.
This cannot possible because the phase space has to be even-dimensional. The solution to this paradox is to notice that the constraint
\begin{displaymath}
    \chi_5=\partial_a \chi_{2,a} + \chi_3=\partial_a \pi_a \ ,
\end{displaymath}
is actually a first-class constraint. Thus, 
we are left with two first-class constraints and six second-class constraints, and the counting 
$    10-(2\cdot 2)-(1\cdot 6)=0$
implies that the theory is fully constrained and no field is dynamical. With this change in the nature of the constraints, the final form of the Hamiltonian is
\begin{align}
    H_f&=\int d^3x\left[ \left(\frac{1}{2} \pi_4^2+ \frac{1}{4} (F_{ab})^2-\pi_a (\partial_a \phi_m)  \right) 
    + (\hat{u}_a-P^T_{ab} A^b ) (\partial_a \pi_4 - \partial_b F^{ba} ) \vphantom{\left( \frac{1}{2} \right)} \right]  \notag \\
    &+ u_1 \pi_5 +u_5 \partial_a \pi_a \ . \label{finalhamiltonian}
\end{align}

\subsection{Gauge symmetry and gauge fixing}

In the previous section we have found two first-class constraints. 
As this type of constraints are related to gauge symmetries, it is worth checking if we have more gauge transformations 
than~\eqref{transformgauge}, inherited from the action in five dimensions. To compute the generators of the gauge transformation, 
we can use Castellani's algorithm~\cite{Castellani} (see also section 5 of~\cite{Pons}). The method states that $G[\alpha]$ is a generator of the gauge transformations if it takes the form
\begin{equation}
    G[\alpha]=G_0  \alpha + G_1 \partial_t \alpha+G_2 \partial_t^2 \alpha + \dots =\sum_{i=1}^N G_i \frac{d^i\alpha}{d t^i} \ ,
\end{equation}
where $G_N$ is a primary first-class constraint, and the subsequent $G$'s fulfil
\begin{equation}
    G_{n-1}=\{G_n , H_f\} + (\text{primary first-class constraints}) \ .
\end{equation}
The process stops when we find a $G_0$ such that $\{G_0 , H_f\}$ is equal only to primary first-class constraints.
In the case we are interested in, this generator takes the form
\begin{equation}
    G[\alpha]=\int d^3 x \left[ \alpha \, \partial_a \pi_a + (\partial_t \alpha) \, \pi_5 \right] \ .
\end{equation}
This is exactly the generator associated with the gauge transformation~\eqref{transformgauge}. 

Now that we are sure that there are no more gauge symmetries, we are in a position to fix the gauge freedom of the theory. As the generator $G[\alpha]$ that we have computed 
is the generator of the gauge transformations, one way to finding a condition $\rho$ that completely fixes the gauge freedom would be to demand that the equation
\begin{equation}
    \{\rho , G[\alpha]\}=0 \ ,
\end{equation}
only has $\alpha=0$ as a solution. Notice that, as we only have one gauge parameter $\alpha$, we can fix all the freedom with one function.

For regular electrodynamics, the most usual gauge-fixing conditions are the Lorenz gauge $\partial_\mu A^\mu=0$, the Coulomb gauge $\partial_a A_a=0$ and the Hamilton gauge $A_t=0$. 
We can check that these three only partially fix the gauge freedom
\begin{align}
    &\{\partial_a A_a+\pi_4 , G[\alpha]\}=\{\partial_a A_a , G[\alpha]\} =\partial_a \partial_a \alpha \ , \\
    &\{A_t , G[\alpha]\}\equiv-\{\phi_m , G[\alpha]\}=\partial_t \alpha \ .
\end{align}
Thus, to completely fix the gauge, we need to choose the combination of the Hamilton gauge with either the Lorenz or the Coulomb gauge.

One may be worried that we are overconstraining our system by imposing two partial gauge conditions (which is what would happen in regular electrodynamics 
if we impose the gauge condition $A_t=A_3=0$), but we can check that it is not the case by computing the Poisson bracket with the Hamiltonian~\eqref{finalhamiltonian}
\begin{align}
    &\{\partial_a A_a , H_f\}=\{\partial_a A_a+\pi_4 , H_f\}=\partial_a \partial_a (u_5 + \phi_m) \ , \\
    &\{\phi_m , H_f\}=-u_1 \ .
\end{align}
As we can see, they transform in conditions for the undetermined functions $u_1$ and $u_5$ and do not lead to additional unwanted constraints.

\subsection{Equations of motion} 
\label{eomqed}

One of the consequences of not having propagating degrees of freedom is that the equations of motion must forbid the propagation of our fields. A quick and heuristic derivation would be to start from the mass-shell condition of the original five-dimensional theory and carrying out the null reduction procedure. Taking the mass-shell condition and demanding that the $s$-component of the momentum has to vanish, we find that the momenta of an excitation in Galilean 
electrodynamics have to fulfil
\begin{equation}
    k_s=k_a=0  \,, \qquad k_t=\text{arbitrary} \,.
\end{equation}
That is, the only physical states allowed are those with zero momentum.~\footnote{Interestingly, the same mass shell condition holds for the different non-relativistic limit of Maxwell equations studied 
in \cite{Palumbo1,Palumbo2}. In particular the authors consider the magnetic limit of the QED Lagrangian up to next-to-leading order thanks to the Hubbard-Stratonovich procedure.} 
To check that this is not an artifact of the null reduction method, we can show that the same happens if we start from the Galilean electrodynamics Lagrangian \eqref{LGED}. Its equations of motion take the form
\begin{align}
    &\partial_a \partial^a \phi_e=0 \ , \label{eom1} \\
    &\partial_a \partial^a \phi_m+\partial_t^2 \phi_e+ \partial_a \partial_t A^a=0 \ , \label{eom2} \\
    &\partial_b \partial^b A^a-\partial_a (\partial_b A^b +\partial_t \phi_e)=0\ . \label{eom3}
\end{align}
If we substitute the gauge fixing $\phi_m=0$, the second equation takes the form
\begin{equation}
    \partial_t (\partial_t \phi_e + \partial_a A^a)=0 \Longrightarrow\partial_a A^a=-\partial_t \phi_e + q(\vec{x}) \ ,
\end{equation}
where $q(\vec{x})$ is an arbitrary function of the spatial variables. However, this relation is actually gauge-dependent and we can use the residual gauge freedom $A^a \to A^a + \partial_a \chi (\vec{x})$ to modify it. The two most interesting choices are setting $q(\vec{x})=0$, which corresponds to the Lorenz gauge, and $\partial_a A^a=0$, which corresponds to the Coulomb gauge.

Let us consider first the Lorenz gauge. If we set $q(\vec{x})=0$, we have $\partial_a A^a +\partial_t \phi_e=0$. Substituting this relation into the third equation, eq.~\eqref{eom3}, we find that the three $A^a$ fulfil the Laplace equation. Therefore, we have
\begin{equation}
    \phi_m=0 \ , \qquad \partial_a \partial^a \phi_e=\partial_a \partial^a A^b=0 \ , \qquad \partial_a A_a +\partial_t \phi_e=0 \ . \label{eomlorenz}
\end{equation}
To further solve these equations we have to make some assumptions about the boundary conditions of our fields. One possibility is to assume that our fields have compact support. This assumption is physically reasonable since 
we are solving electromagnetism in the vacuum. However, it is too restrictive because the Laplace equations does not admit non-trivial solutions with compact support, so all our fields would vanish.
A milder assumption would be to demand compact support for the electric 
and magnetic fields, defined as~\footnote{Another possible boundary condition would be to demand $A_a$ and $\phi_e$ to be bounded. 
One can check that such boundary condition is more restrictive than the one we will be considering.} 
\begin{equation}
    E_a=\pi_a=\partial_a \phi_e \ , \qquad B_a=\epsilon_{abc} \partial_b A_c \ . \label{EandB}
\end{equation}
The fact that $\phi_e$ and $A_a$ fulfil a Laplace equation implies that the electric field is solenoidal, and the magnetic field is irrotational. Thus, the solution compatible with these conditions is
\begin{equation}
    A^a=\partial^a g(\vec{x},t) - \frac{1}{3}x^a \,h(t) \ , \qquad \partial_a \partial^a  g(\vec{x},t)=0 \ , \qquad \phi_e=h(t)\ , \qquad \phi_m=0 \ ,
\end{equation}
where $h(t)$ is a generic function of time and $g(\vec{x},t)$ is a solution to the Laplace equation.

We will now move to the Coulomb gauge. If we set $\partial_a A_a=0$, we have $\partial_t \phi_e=q(\vec{x})$. Applying the Laplacian on both sides, equation~(\ref{eom1}) implies that $q(\vec{x})$ satisfies the Laplace equation. 
If we apply instead a time derivative, we conclude that $\partial_t^2 \phi_e=0$. Therefore, we have the following conditions
\begin{align}
    \phi_m=0 \ , \quad \partial_a A^a =0 \ , \quad \phi_e =r(\vec{x}) + t \, q(\vec{x}) \ , \notag \\
    \partial_a \partial^a A_b =\partial_b q (\vec{x})\ , \quad  \partial_a \partial^a r(\vec{x})  =\partial_a \partial^a q(\vec{x})=0\ .
\end{align}
If we assume that the electric and magnetic fields have compact support, the solution to the above equations is
\begin{equation}
    A^a=\partial^a g(\vec{x},t) \ , \qquad \phi_e=\alpha + \beta t \ , \qquad \phi_m=0 \ ,
\end{equation}
where $\alpha$ and $\beta$ are constants and $g(\vec{x},t) $ is again a solution to the Laplace equation.

In contrast with the case of relativistic quantum electrodynamics, there are no plane waves after we impose any of the two gauge fixing conditions. 
This happens because the equations of motion we find are elliptic instead of hyperbolic. As we will discuss in the following section, this has important consequences 
for the quantization of the theory.  

\section{A propagator paradox}

When we proceed to compute the propagators associated to the field content of Galilean electrodynamics we find a striking paradox. 
On the one hand, a useful method to compute propagators is to invert the quadratic part of the action in Fourier space. 
This is what was done for Galilean electrodynamics in~\cite{Chapman:2020vtn} for 2+1 dimensions and  in~\cite{Banerjee:2022uqj} for 3+1 dimensions. 
In both cases, the authors found non-trivial propagators. On the other hand, the analysis of the equations of motion from the previous section indicates 
that there is no oscillatory dependence on the spatial directions. If we follow the usual computations of relativistic quantum electrodynamics
(see e.g. chapter 7 of~\cite{Field_Quantization}), we see that these oscillatory modes are what leads to a non-trivial propagator in canonical quantization. 
Thus, the propagators of these fields should be trivial.

The solution to this paradox is to remind ourselves that this is a constrained theory. 
This was addressed both in references~\cite{Chapman:2020vtn} and~\cite{Banerjee:2022uqj} by the Faddeev-Popov method. However, there is an important issue with the approach 
of these authors, which is that the Faddeev-Popov method only works for first-class constraints.~\footnote{Even in the hypothetical case that the Faddeev-Popov trick would work in this situation, 
the authors of these two articles only fix Lorenz gauge. This, as we have shown above, does not fix all the gauge degrees of freedom.} 
The approach to deal with a system with both first-class and second-class constraints
requires to us impose every constraint that we have through delta functions in the path integral~\cite{Fradkin,Senjanovic}. 
In particular, given a four-dimensional system with first-class constraints $\phi_i$, gauge conditions for the first-class constraints $\rho_i$, second-class constraints $\chi_i$, 
and Hamiltonian $H$, regardless of the nature of the constraints, the path integral is given by \cite{Senjanovic}
\begin{align}
    Z=&\int \mathcal{D}\psi_a \mathcal{D}\pi_b \prod |\det(\{\phi_a , \rho_b\})| \delta (\phi_a)\delta (\rho_a) \prod|\det(\{\chi_a , \chi_b\})|^{1/2} \delta (\chi_a) \notag \\
    &\exp\left[ i\int{(\pi_a \psi^a -H) \,d^4x} \right] \ .
\end{align}
Notice that the expression does not care if we include the constraints in the Hamiltonian~$H$, as the terms proportional to the Lagrange multipliers vanish due to the delta functions. 

When we apply the above formula to the Hamiltonian \eqref{HGED} with the constraints computed in section~\ref{constraints} and impose trhe Coulomb gauge, we get
\begin{align}
    Z =&\int \mathcal{D}\mu\,  \delta (\phi_m) \delta (\pi_5) \delta (\partial_a \pi_a) \delta (\partial_a A_a)   \delta^{(3)}(\pi_a - \partial_a \phi_e) \delta^{(3)}(\partial_a \pi_4 - \partial_b F^{ba}) \notag \\
    &\exp\left[ i\int{(\pi_a \partial_tA_a + \pi_4 \partial_t\phi_e + \pi_5 \partial_t\phi_m -H_{GED} - J_a A_a - J_e \phi_e - J_m \phi_m) \,d^4x} \right] \ ,
\end{align}
where $\mathcal{D} \mu$ stands for the integration measure together with the determinant of the constraints. Notice that we have included sources for all the fields in order 
to compute correlation functions later. Integrating over $\pi_a$, $\pi_5$ and $\phi_m$ by means of the delta functions we find
\begin{align}
    Z =&\int \mathcal{D}\mu'\,  \delta (\partial_a \partial_a \phi_e) \delta (\partial_a A_a)   \delta^{(3)}(\partial_a \pi_4 - \partial_b F^{ba}) \notag \\
    &\exp\left[ i\int{ \left( \partial_a \phi_e \partial_tA_a + \pi_4 \partial_t\phi_e -\frac{1}{2} \pi_4^2 - \frac{1}{4} (F_{ab})^2 - J_a A_a - J_e \phi_e \right) \,d^4x} \right] \ ,
\end{align}
If we recall now the split of the constraint $\chi_{4,a}$ that we discussed in \eqref{splitconstraint}, we need to demand that $\pi_4$ has to be a solution of the Laplace equation. 
If we impose that $\pi_4$ is boundded at infinity, the only possibility is for it to be an arbitrary function of time, $p(t)$. If we demand that the electric and magnetic fields, defined 
in equation~\eqref{EandB}, have compact support, a similar reasoning applies to $\phi_e$, which we will denote as $q(t)$, while  $A_a$ has to be the gradient of the solution to a Laplace equation, 
which we will denote by $\partial_a g(\vec{x},t)$. Therefore, the delta functions become
\begin{align}
    Z =&\int \mathcal{D}\mu^{\prime \prime}\,  \delta [\phi_e - q(t)] \delta [ A_a - \partial_a g(\vec{x},t)]   \delta^{(3)}[\pi_4 - p(t)] \notag \\
    &\exp\left[ i\int{ \left( \partial_a \phi_e \partial_tA_a + \pi_4 \partial_t\phi_e -\frac{1}{2} \pi_4^2 - \frac{1}{4} (F_{ab})^2 - J_a A_a - J_e \phi_e \right) \,d^4x} \right] \nonumber \\ 
    &=C\int \mathcal{D}g \mathcal{D} p \mathcal{D}q \, \exp\left[ i\int{ \left[ V\left( p \partial_t q -\frac{p^2}{2} \right) - j_e q \right] \,dt -i \int{ J_a \partial_a g}\,d^4x } \right] \ ,
\end{align}
where $C$ is a global constant of no relevance, $V$ is the volume of the space, and $j_e$ is the current $J_e$ integrated over the spatial directions. 
As the path integral is linear in $g$, we can integrate over it, which provides a delta function for $\partial^a J_a$. Once this is done, the path integral reduces to the one 
of a free quantum-mechanical particle with a source term, 
\be
Z \sim \delta (\partial^a J_a) \exp \left( \frac{i}{2V} \int{dt \, dt' j_e (t) \frac{1}{\partial_t^2} j_e (t')} \right) \ .
\ee
From this path integral it is immediate to see that the only non-vanishing two-point function of the theory would be the one involving two fields $\phi_e$, which will take the following form in Fourier space, 
\begin{equation}
    \langle \phi_e \phi_e\rangle \sim \frac{\delta^3 (\vec{k})}{\omega^2} \ .
\end{equation}
In position space, this two-point function takes the form $\langle \phi_e \phi_e\rangle \sim t^2$, which perfectly matches the proposal in \cite{Bagchi:2014ysa}, 
as $\phi_e$ has classical conformal dimension $\Delta=1$.

We can repeat the computations replacing the Coulomb gauge $\partial_a A_a=0$ by the Lorenz gauge $\partial_a A_a + \partial_t \phi_e=0$. 
The computation follows along the same steps as above. The difference appears when imposing the boundary conditions to $A^a$. Demanding the Lorenz gauge condition together with compact support 
of the magnetic field, the constraint $\delta [ A_a - \partial_a g(\vec{x},t)]$ has to be replaced by $\delta [ A_a - \partial_a g(\vec{x},t)+\frac{1}{3} x_a \pi_4]$, so the path integral takes the form
\begin{align}
    Z &\sim\int \mathcal{D}g \mathcal{D} p \mathcal{D}q \, \exp\left[ i\int{ \left[ \left( p \partial_t q -\frac{p^2}{2} \right) - J_a \left(\partial_a g - \frac{1}{3} x_a p \right) - J_e q \right] \,d^4x} \right]  \\
    &\sim \delta (\partial^a J_a) \exp \left[ \frac{i}{18V} \int{k^2 dt} - \frac{i}{2V} \int{dt \, dt' \left( j_e (t)+\frac{\partial_t k(t)}{3}\right) \frac{1}{\partial_t^2} \left( j_e (t') +\frac{\partial_tk(t')}{3}\right)} \right] \ , \nonumber
\end{align}
where $k=\int{ J_a x^a d^3 x}$ is the dipole momentum associated to $J_a$.
From the path integral we conclude  that the two-point function $\langle \phi_e \phi_e\rangle$ has the same form as in the Coulomb gauge. 
However, in this case the two-point function $\langle \phi_e A^a\rangle$ is non-vanishing and proportional to~$t x^a$. 
This does not seem to match the proposal in \cite{Bagchi:2014ysa}, which indicates that the field configuration for $A^a$ is not invariant under translations or Galilean boosts.

%%%%%%%%%%%%%%%%%%%%%%%%%%%%%%%%%%%%%%%%%%%%%%%%%%%%%%%%%%%%%%%%%%%%%%%%%%%%%%%%%%%%%%%%%%%%%%%%%%%%
%%%%%%%%%%%%%%%%%%%%%%%%%%%%%%%%%%%%%%%%%%%%%%%%%%%%%%%%%%%%%%%%%%%%%%%%%%%%%%%%%%%%%%%%%%%%%%%%%%%%

\section{Conclusion}

In this article we have studied the dynamical structure of the Galilean electrodynamics Lagrangian and its gauge symmetry. We have found that the theory has two first-class constraints 
and six second-class constraints. As a consequence, the theory is fully constrained and has zero degrees of freedom. We have also studied the implications that this has at the level 
of the equations of motion and the path integral. At the level of the equations of motion, we have found that all the fields fulfil the Laplace equation when imposing both the $\phi_m=0$ gauge and the Lorenz gauge. 
This means that these fields do not carry any oscillatory modes and, thus, the propagator of these fields cannot resemble the relativistic photon propagator. To find the actual propagators 
of the theory, we have computed the path integral using the methods for constrained theories developed in \cite{Fradkin,Senjanovic}. We have found that the path integral contains a single zero mode. 
This differs from the result for the propagators in \cite{Chapman:2020vtn} and \cite{Banerjee:2022uqj}, where the constraints of the theory were not properly taken into account. 

We should stress that along our discussion of the path integral we have assumed that all our physical field have compact support, 
which heavily restricts the allowed solutions to the Laplace equations we have to solve. Nevertheless, it would be interesting to analyse the theory with more general boundary conditions. 
We must note that the fact that the theory depends on the boundary conditions at infinity does not mean that the theory is topological. 
The absence of propagating local degrees of freedom is a necessary condition to have a topological quantum field theory, but not sufficient. 
The fact that the equations of motion involve the spatial Laplacian, and thus carry an explicit dependence on the metric, is enough to claim that it is not topological.

A very interesting generalization of this work would be to extend the analysis to Galilean Yang-Mills, due to its connection to non-relativistic GGK/GYM holography 
(see~\cite{Fontanella:2026gaq} for a recent review on the topic). As the quadratic terms in Galilean electrodynamics and Galilean Yang-Mills are the same, we might be tempted 
to say that the latter theory would also be fully constrained. This seems to align with the results obtained from string theory, where it was found that the Lagrangian of fluctuations 
around the folded BMN-like string are given by free fields in AdS$_2$ space \cite{deLeeuw:2024ijj}. Nevertheless, a detailed computation is needed as new structures emerge. 
For example, ghosts can be safely ignored for Galilean electrodynamics in the same way as in relativistic quantum electrodynamics, but this is not the case for Yang-Mills. In addition, 
cubic and quartic terms in the fields will appear in the constraints that may alter the analysis we have performed here. We should also note that quantization of Galilean Yang-Mills 
has been studied in~\cite{Bagchi:2022twx}, but this article borrows the propagators from \cite{Chapman:2020vtn}, and thus suffers from the same issues. 

%%%%%%%%%%%%%%%%%%%%%%%%%%%%%%%%%%%%%%%%%%%%%%%%%%%%%%%%%%%%%%%%%%
%%%%%%%%%%%%%%%%%%%%%%%%%%%%%%%%%%%%%%%%%%%%%%%%%%%%%%%%%%%%%%%%%%

\vspace{8mm}

\centerline{\bf Acknowledgments}

\vspace{2mm}

\no
The work of R.~H. is supported by grant PID2023-149834NB-I00. J.~M.~N.~G. wants to thank A.~Fontanella for useful discussion on the topic and comments on the manuscript.

%%%%%%%%%%%%%%%%%%%%%%%%%%%%%%%%%%%%%%%%%%%%%%%%%%%%%%%%%%%%%%%%%%
%%%%%%%%%%%%%%%%%%%%%%%%%%%%%%%%%%%%%%%%%%%%%%%%%%%%%%%%%%%%%%%%%%

\end{document}